\documentclass{optica-article}
\journal{oe} 

\usepackage{ulem}
\usepackage{hyperref}
\usepackage{soul}

\begin{document}

\title{Bandgap fluctuations and robustness in two-dimensional hyperuniform dielectric materials.}
\author{Luis S. Froufe-Pérez\authormark{1,3,$\dagger$}, Geoffroy J. Aubry\authormark{1,2,4,$\dagger$}, Frank Scheffold\authormark{1,5}, and Sofia Magkiriadou\authormark{1,6}}

\address{\authormark{1}Department of Physics, University of Fribourg, 1700 Fribourg, Switzerland\\
\authormark{2}Université Côte d’Azur, CNRS, Institut de Physique de Nice -- INPHYNI, France\\
\authormark{3}luis.froufe@unifr.ch\\
\authormark{4}geoffroy.aubry@cnrs.fr\\ 
\authormark{5}frank.scheffold@unifr.ch\\
\authormark{6}sofia.magkiriadou@unifr.ch\\
\authormark{$\dagger$}Equal first authorship
}

\begin{abstract*}
We numerically study the statistical fluctuations of photonic band gaps in ensembles of stealthy hyperuniform disordered patterns. We find that at low stealthiness, where correlations are weak, band gaps of different system realizations appear over a wide frequency range, are narrow, and generally do not overlap. Interestingly, above a critical value of stealthiness $\chi \gtrsim 0.35$, the bandgaps become large and overlap significantly from realization to realization, while a second gap appears. These observations extend our understanding of photonic bandgaps in disordered systems and provide information on the robustness of gaps in practical applications.
\end{abstract*}


\section{Introduction}
The propagation of light depends on a material’s internal structure. If inhomogeneities on length scales of the order of the wavelength of light are present, interference effects can lead to a variety of phenomena that affect transport. Among these, one of the most celebrated is the existence of photonic bandgaps \cite{Yablonovitch1987,Joannopoulos2008}. Photonic band gaps in crystalline dielectric materials have been extensively discussed, since there the symmetries of the lattices can directly lead to forbidden states for light of certain frequencies. However, according to recent studies, order is not required for the formation of photonic band gaps, as they have also been observed in disordered materials~\cite{Florescu2009}. In particular, some types of disordered materials can exhibit isotropic, complete, and wide photonic bandgaps similar to those of crystal structures made of the same material~\cite{Florescu2009, Klatt2021}. A class of those are known as disordered hyperuniform materials \cite{Torquato2003, Leseur2016, ricouvier2017optimizing, Froufe-Perez2016, Froufe-Perez2017, wilken2020hyperuniform}, characterized by the lack of long-range density fluctuations~\cite{Froufe-Perez2017}. The degree of order in disordered stealthy hyperuniform materials can be quantified by the stealthiness parameter, $\chi$ \cite{torquato2015ensemble}: In two dimensions, when $\chi = 0$, the system is random leading to large density fluctuations; as $\chi$ approaches $0.5$ the system becomes more and more uniform while remaining isotropic. For $\chi \gtrsim 0.5$ isotropy is gradually lost and structures become crystalline~\cite{torquato2015ensemble}. 

The understanding of photonic bandgaps in crystals is greatly facilitated by the fact that they are uniquely defined: given (i) a unit cell, (ii) the fundamental blocks that decorate the lattice points (e.g. cylinders or spheres), and (iii) the type of connection between them (e.g. whether they are connected by walls or not) as well as (iv) the material properties (e.g. the dielectric constant), the system can be reconstructed with no ambiguity and the corresponding optical properties can be uniquely determined. However, the same is not true for disordered materials.
Hyperuniform structures have no primitive vectors, and even if the value of $\chi$ is specified, there are many possible realizations of a point pattern~\cite{torquato2015ensemble}. Therefore, to arrive at a comprehensive understanding of photonic bandgaps in disordered hyperuniform materials, we need to consider a statistical ensemble of them~\cite{Klatt2022PNAS}. This naturally gives rise to the following question: given a value of stealthiness, are the photonic bandgaps of disordered hyperuniform materials uniquely defined, or is there statistical variability in their properties, and if so, how strong is it?

This question is interesting both from a fundamental as well as from a technological point of view \cite{wiersma2013disordered}. Fundamentally, identifying the salient features that determine the presence or absence of photonic bandgaps in disordered systems will advance our understanding of their origin~\cite{Monsarrat2022}. Technologically, disordered photonic materials have already been shown to be relevant for various applications, such as waveguides with large bending angles, light harvesting, or quantum applications \cite{Man2013, Ma2016, florescu2013optical, tavakoli2022over, granchi2022near}. If the photonic bandgaps of a disordered material turn out to be robust to statistical variability in the material's density fluctuations, this would suggest that devices based on these materials could require less precision in their manufacture compared to their crystalline counterparts. Therefore, addressing this question can give us information about the allowed tolerance in practical applications.

Here we investigate this question by examining a large number of different realizations of disordered two-dimensional dielectrics. For this purpose, we create point patterns for different values of $\chi$, decorate them with dielectric disks, and compute the corresponding band structures numerically. Technical details of the numerical calculations are described in detail in one of our early works~\cite{Froufe-Perez2016}. 
We consider TM modes (electric field perpendicular to the plane) and we calculate the DOS for $\chi$ from 0.1 up to 0.5. We then look at the distributions of the bandgap central frequencies and of their widths, and we show how these distributions change with $\chi$. We find that we can always define a bandgap. For small values of $\chi \lesssim 0.2$, however, the gaps are very narrow. As $\chi$ increases, the gap width also increases. However, for the same value of $\chi$, initially there is still a large variability in the gap widths and central frequencies, so that on average there is no gap.
Interestingly, we observe a transition at $\chi \sim 0.4$ where the gaps become wide and robust and also a second gap appears. 

\begin{figure}
    \centering
    \includegraphics{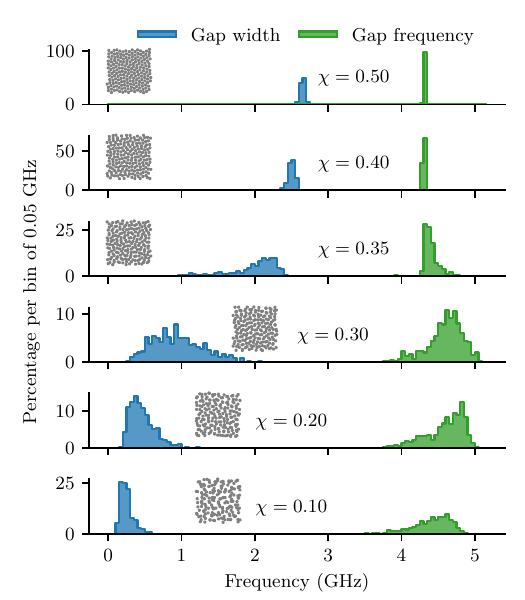}
    \caption{Histograms of the widths (left, in blue) and of the mean frequencies (right, in green) of the largest gap found in each single sample. The bin width is 0.05 GHz and each histogram is normalized so that the sum of the bar heights amounts to 100$\%$.
    An example of a point pattern is shown for each $\chi$-value.}
    \label{fig:histograms}
\end{figure}

\section{Simulation Results and Discussion}

We generate 500 different 2D point patterns for each value of $\chi = 0.10$ up to 0.50, in steps of 0.05 as described in ~\cite{Froufe-Perez2016}. All point patterns have the same number density of $\rho = 0.32$~cm$^{-2}$ and around 200 scatterers. These scatterers are dielectric cylinders of radius 3~mm and dielectric permittivity $\varepsilon=37$ (refractive index $n=\sqrt{\varepsilon}=6.08$ in air) to match the setup used in ref.~\cite{Aubry2020}. Each point pattern was generated using periodic boundary conditions, and we therefore can use the supercell method to compute the states~\cite{Joannopoulos2008}.
For any $\vec{k}$ in the reciprocal space, we compute the bands for TM modes by direct diagonalization of the Hamiltonian of the system in the planewave basis~\cite{Joannopoulos2008}, i.e, 
\begin{align}
\sum_{\vec{G'}}\left|\vec{k}+\vec{G'}\right|\left|\vec{k}+\vec{G}\right|\tilde{\varepsilon}_{\vec{G}\vec{G'}}^{-1}h_{\vec{G'}}=\left[\frac{\omega(\vec{k})}{c}\right]^{2}h_{\vec{G}}
\end{align}
where $\omega(\vec{k})$ are eigenfrequencies, $c$ is the speed of light in vacuum, $h_{\vec{G}}$ is the $\vec{G}$-component of the expansion of the magnetic field in planewaves, and $\vec{G}$ or $\vec{G'}$ are reciprocal lattice vectors.
Here $\tilde{\varepsilon}_{\vec{G}\vec{G'}}^{-1}$ is the element $(\vec{G},\vec{G'})$ of the inverse of the matrix whose elements $\tilde{\varepsilon}_{\vec{G}\vec{G'}}$ are the spatial Fourier transform of the permittivity taken at $\vec{G}-\vec{G'}$.
We calculate each band on 12 points $\vec{k}$ in the reciprocal lattice along the $\Gamma-X-M-\Gamma$ path using 8000 points $\vec{G}$ in the reciprocal lattice.

Figure~\ref{fig:histograms} shows the distributions of the bandgap widths and of their mean frequencies for the 500 system realizations at different values of $\chi$.
For each system realization, we identify the largest gap as follows. First, we take the difference between the highest value of the frequency of each band and the lowest value of the band above it.
This quantity might be negative (indicating no gap), or positive (indicating a gap).
We then take the maximum value of these differences and call it the \textit{largest gap} for each sample.
For all our samples, the largest of these differences is always positive, and the central frequency of this largest gap is always within the range of 4 - 5 GHz; this is the first gap.
At small values, $0.1 \leq \chi \leq 0.2 $, these bandgaps tend to be narrow (width $<$ 1 GHz), and their mean frequencies are broadly distributed. We interpret the narrow band gaps at small $\chi$s as a consequence of statistical fluctuations in the finite size systems. To keep our analysis consistent, we report all gap values obtained using the same scheme.

\begin{figure}[h]
    \centering\includegraphics{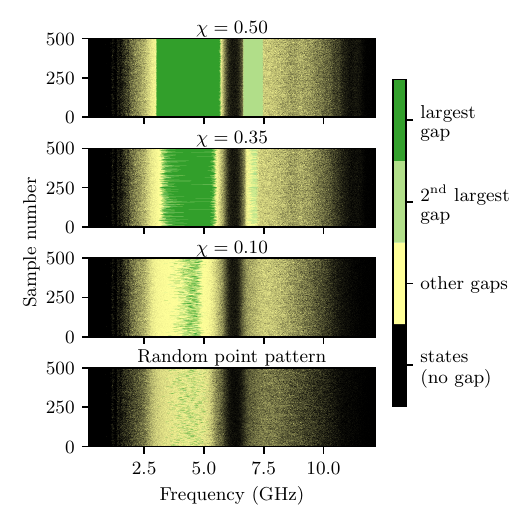}
    \caption{Map of the locations of states and gaps, in frequency, for the 500 system realizations of random point patterns, and of samples at three different values of $\chi$: 0.10, 0.35 and 0.50.
    The largest gap is shown in dark green, the second largest gap in light green.
    Any other gap is shown in yellow, and black denotes states.
    Note that even though the yellow zone might appear larger than the dark green region in the $\chi=0.10$ panel, there are some states in-between.}
    \label{fig:gap_overlap}
\end{figure}

In Fig.~\ref{fig:gap_overlap} we represent the largest gap of each sample on the frequency axis. 
Each line represents a sample, and the largest gap is shown in dark green.
In this representation, we see that for $\chi=0.10$, the largest gaps do not overlap from sample to sample.
As $\chi$ increases, but still below 0.35, the distribution of gap widths shifts to higher values and broadens (see blue histograms in Fig.~\ref{fig:histograms}): bandgaps open up, however there is variability in their widths. Similarly, there is still variability in the gaps' mean frequencies for intermediate $\chi$-values (see green histograms in Fig.~\ref{fig:histograms}).
It is only when $\chi$ reaches $\chi \geq 0.4$ that, remarkably, both the mean gap frequencies and their widths collapse into narrow distributions: the bandgaps are very similar among different realizations of stealthy hyperuniform patterns.
This behavior can be seen clearly in Fig.~\ref{fig:gap_overlap} where the dark green region shows that all gaps overlap with a lower and an upper bandedge that fluctuate much less in the $\chi=0.50$ case than in the $\chi=0.35$ case. 

In Fig.~\ref{fig:gap_overlap} we also highlight the gap with the second largest width in light green.
For $\chi=0.10$, the second largest gap central frequency is distributed over the same frequency range as the largest gap which results in a difficult distinction in the colormap.
The picture is very different in the $\chi=0.50$ case where all the second largest gaps for the different samples have very similar central frequencies and widths.
This second largest gap corresponds, for the high-$\chi$-values, to the second bandgap reported experimentally in ~\cite{Aubry2020}.

\begin{figure}[h!]
    \centering
    \includegraphics{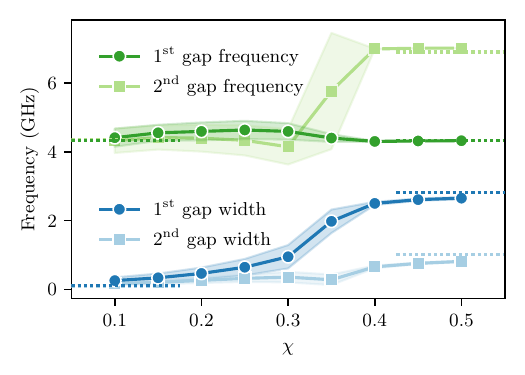}
    \caption{Average value of the width (bottom two curves in blue) and of the frequency (top two curves in green) of the two bandgaps (first gap shown in dark colour, circles; second gap shown in light colour, squares). The shaded area around each curve denotes the standard deviation.
    The dotted lines show the corresponding values for the two first gaps of a random point pattern on the left, and of the triangular lattice on the right.}
    \label{fig:widthAndFreq}
\end{figure}

To further quantify the statistics of the two largest gaps in these systems, we plot in Fig.~\ref{fig:widthAndFreq}, as a function of $\chi$, the central frequencies (in green) and widths (in blue) of the first and second bandgap averaged over the 500 numerical samples.
The width of the error bars is given by the standard deviation of the distribution.
Moreover, the dotted horizontal lines on the left show the central frequencies and widths of the two first TM gaps of a random point pattern, and those on the right of a triangular lattice, both with the same density of cylinders as the hyperuniform samples.
We see that, in all cases, the standard deviations vanish for $\chi\geq0.40$.
For these high $\chi$-values, the averages of the widths continue to increase toward the widths of the gaps in the triangular lattice, in contrast to the central frequencies which already converge at $\chi=0.40$. 

\begin{figure}[h!]
    \centering
    \includegraphics{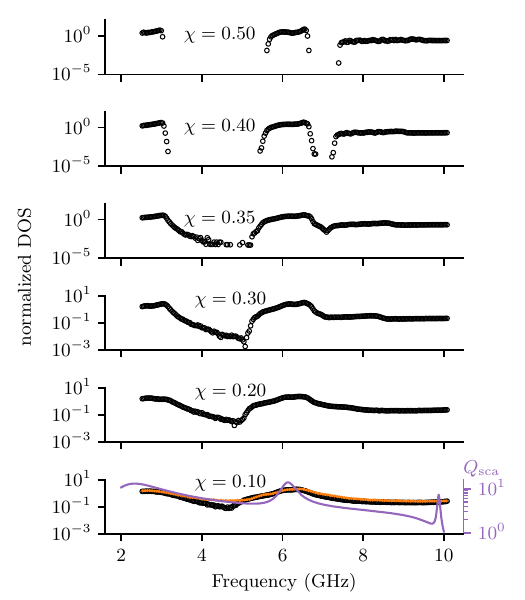}
    \caption{Normalized density of states (nDOS) obtained by taking the average over the band structure calculated numerically for 500 system realizations at each value of $\chi$.
    In the lower panel ($\chi = 0.10$), the orange symbols show the nDOS for a completely uncorrelated system, while the thin solid purple line is the scattering efficiency $Q_\mathrm{sca}$ of a single scatterer. Note that the nDOS for the uncorrelated system and for $\chi = 0.1$ have significant overlap.}
    \label{fig:numericalDOS}
\end{figure}

Figure~\ref{fig:numericalDOS} shows the normalized density of states (nDOS) of these stealthy hyperuniform samples for each value of $\chi$. Normalization is performed by dividing the histogram of eigenfrequencies by a linear density of states that would be obtained in a homogeneous medium.
The nDOS was calculated with the same data as used for the bandgap statistics.
The nDOS always has a minimum; however, there is a drastic decrease in the value of this minimum above $\chi \geq 0.4$, as it reduces to 0 over a large range of frequencies. Signatures of the gaps are well visible for $\chi\leq 0.40$. Below these values, the nDOS shows a minimum as a precursor for the first gap, and a shoulder as a precursor for the second gap.

The detected first and second largest gap central frequencies are similar for low $\chi$-values (see green curves in Fig.~\ref{fig:widthAndFreq}).
As soon as the second gap at $\sim 7$~GHz becomes more robust, i.e. for $\chi=0.35$, the likelihood to find the second largest gap around 7~GHz increases, as can be seen from the enormous broadening of the standard deviation of the light green curve from of Fig.~\ref{fig:widthAndFreq}. For $\chi\geq0.40$, both gaps are well visible in the nDOS: this correlates well with the bandgap statistics for high $\chi$-values presented in Fig.~\ref{fig:histograms} and ~\ref{fig:gap_overlap}.

The opening of the bandgap of the ensemble can be understood if we consider the distributions of the gaps' mean frequencies and widths, shown in Fig.~\ref{fig:gap_overlap} (see also Fig.~\ref{fig:histograms}) as well as the scattering properties of individual cylinders. For the fully ordered case of photonic crystals, Vynck and coworkers proposed that both gaps are created by the electric dipole resonance of the cylinders, but the magnetic dipole resonance occurs in the middle and separates the gap into two parts by a conduction band \cite{vynck2009all}.
For disordered structures, at low $\chi$, even though every configuration has a bandgap, these gaps are narrow and vary in frequency; thus, they do not have much overlap and, on average, the nDOS at the corresponding frequency range is non-zero (see the $\chi=0.10$ case in Fig.~\ref{fig:gap_overlap}).
The picture is different at $\chi \geq 0.4$, where the mean frequencies of the gaps do not vary as much from realization to realization, while at the same time the gaps are much wider; as a result, they have greater overlap, as shown in Fig.~\ref{fig:gap_overlap} for the case of $\chi = 0.5$. Consequently, the nDOS of the ensemble remains 0 over a large frequency range of 3 GHz – 5.5 GHz, which corresponds to the mean location and width of the gaps shown in Fig.~\ref{fig:histograms} and to the large gaps shown in Fig.~\ref{fig:gap_overlap} and \ref{fig:numericalDOS}.

\section{Conclusions and Outlook}

Our results offer insight towards the understanding of the ingredients required for the appearance of bandgaps in disordered systems, and they emphasize the importance of correlations for the formation and robustness of photonic bandgaps. We find that for finite pieces of 2D materials we can always find a bandgap. However, when $\chi$ is small, it is difficult to predict the location and width of the bandgap for each realization. The gaps can be wider or very narrow, and their central frequencies can be found over a wide range of values. On the other hand, for $\chi \geq 0.4$, the location and width of the bandgap are robust features of the system regardless of the details of the scatterers' distribution. Thus, it is likely that these features are also robust to perturbations or imperfections, when it comes to practical realization of bandgap materials. 

Of course many questions remain. To complete the picture, it would be interesting to analyze band gaps in both TM and TE polarizations, as well as to investigate the effect of the refractive index contrast on the critical value of $\chi$. However, we assume that differences in system parameters do not qualitatively change the picture. Interestingly, we observe a sharp change in the bandgap statistics at $\chi \gtrsim 0.35$ which raises the question whether our observations could be described by a (critical) phase transition---a topic for future work. Finally, in the present work we have not discussed the influence of system size, as all of our systems were composed of around 200 scatterers. Would our observations be relevant for larger systems? An interesting contribution was made recently in this direction, where the authors investigated the behavior of photonic bandgaps in disordered hyperuniform systems of increasing system size and extrapolated to the thermodynamic limit, finding, similarly to our results, a critical $\chi$ value above which bandgaps are robust~\cite{Klatt2022PNAS}. This is also compatible with the results presented in \cite{Froufe-Perez2017} on transport statistics in equivalent systems of different sizes. Considering a large system as the sum of many smaller ones, one might hypothesize that the DOS of the entire system would resemble the average of the nDOS of its smaller parts. If these have overlapping gaps, then it might be reasonable to expect that these gaps will be present for the entire system, whereas if the smaller sub-systems only have non-overlapping gaps, then perhaps no gap would appear overall. If true, this approach would provide a new method to study larger disordered hyperuniform systems, a task that is at present limited by the computational complexity that comes with lack of periodicity. 

\begin{backmatter}
\bmsection{Funding} The Swiss National Science Foundation financially supported this work through the National Center of Competence in Research Bio-Inspired Materials, No.~182881 and through projects No.~169074, No.~197146, and No.~188494. We acknowledge the OPAL infrastructure and the Center for High-Performance Computing from the Université Côte d'Azur (ANR-15-IDEX-01, Idex UCAJEDI).

\bmsection{Acknowledgments} Our comprehensive re-examination of the band structure of two-dimensional disordered ~\cite{Aubry2020} dielectrics was inspired by the recent research conducted by Klatt, Steinhard, and Torquato~\cite{Klatt2022PNAS}. We express gratitude to Salvatore Torquato and Paul Steinhardt for insightful conversations and valuable feedback on our manuscript. We thank Kevin Vynck for bringing to our attention their discussion about the origin of band gaps in two-dimensional photonic crystals constructed from dielectric cylinders~\cite{vynck2009all}.

\bmsection{Disclosures} The authors declare no conflicts of interest.

\bmsection{Data availability} Data underlying the results presented in this paper will be made available upon publication via a public repository \cite{data_repo}.
\end{backmatter}

\bibliography{sharedBiblio}

\end{document}